\documentclass{iopart}
\usepackage{hyperref}
\usepackage{amssymb}
\usepackage{graphicx}  
\begin{document}

\title[Search for GWs associated with GRB 050915a.]{Search for gravitational waves associated with GRB~050915a using the Virgo detector}

\author{F.Acernese $^{7,9}$,
M.Alshourbagy$^{15,16}$,
P.Amico$^{13,14}$,
F.Antonucci$^{19}$,
S.Aoudia$^{10}$,
K.G. Arun$^{11}$,
P.Astone$^{19}$,
S.Avino$^{7,8}$,
L.Baggio$^1$,
G.Ballardin$^2$
% ...
}

\address{$^1$Laboratoire d'Annecy-le-Vieux de Physique des Particules (LAPP),  IN2P3/CNRS, Universit\'e de Savoie, Annecy-le-Vieux, France}
\address{$^2$European Gravitational Observatory (EGO), Cascina (Pi), Italia.}
% ...
\address{$^7$ INFN, sezione di Napoli, Italia.}
\address{$^8$ Universit\`a di Napoli "Federico II" Complesso Universitario di Monte S.Angelo, Italia.}
\address{$^9$ Universit\`a di Salerno, Fisciano (Sa), Italia.}
\address{$^{10}$Departement Artemis -- Observatoire de la C\^ote d'Azur, BP 4229 06304 Nice, Cedex 4, France.}
\address{$^{11}$LAL, Univ Paris-Sud, IN2P3/CNRS, Orsay, France.}
% ...
\address{$^{13}$INFN, Sezione di Perugia, Italia.}
\address{$^{14}$Universit\`a di Perugia, Perugia, Italia.}
\address{$^{15}$INFN, Sezione di Pisa, Italia.}
\address{$^{16}$ Universit\`a di Pisa, Pisa, Italia.}
% ...
\address{$^{19}$INFN, Sezione di Roma, Italia.}
% ...

\begin{abstract}
In the framework of the expected association between gamma-ray bursts and gravitational waves, we present results of an analysis aimed to search for a burst of gravitational waves in coincidence with gamma-ray burst 050915a. This was a long duration gamma-ray burst detected by \textit{Swift} during September 2005, when the Virgo gravitational wave detector was engaged in a commissioning run during which the best sensitivity attained in 2005 was exhibited. This offered the opportunity for Virgo's first search for a gravitational wave signal in coincidence with a gamma-ray burst. The result of our study is a set of strain amplitude upper-limits, based on the loudest event approach, for different but quite general types of burst signal waveforms. The best upper-limit strain amplitudes we obtain are $h_{rss}={\cal O}(10^{-20})$~Hz$^{-1/2}$ around $\sim 200-1500$~Hz. These upper-limits allow us to evaluate the level up to which Virgo, when reaching nominal sensitivity, will be able to constrain the gravitational wave output associated with a long burst. Moreover, the analysis here presented plays the role of a prototype, crucial in defining a methodology for gamma-ray burst triggered searches with Virgo and opening the way for future joint analyses with LIGO.	    
\end{abstract}


\maketitle

\section{Introduction}
Gamma-Ray Bursts (GRBs) are intense flashes of $\gamma$-ray (and X-ray) photons, lasting from few milliseconds to several minutes, followed by a fainter and fading emission at longer wavelengths called the ``afterglow'' \cite{PiranReview2005,MeszarosReview2006}.  
GRBs are detected at a rate of about one per day, from random directions in the sky. 
They fall into two apparently distinct categories, namely short-duration (nominally, less than 2 s), hard-spectrum bursts (short GRBs) and long-duration (greater than 2 s), soft-spectrum bursts (long GRBs) \cite{Dezalay1992,Kouvelioutou1993,soma1998,Preece2000}. Of course, this separation is not strict and the two populations do overlap, but such a distinction has suggested two different types of progenitors. Progenitors of long GRBs are thought to be massive, low-metallicity stars exploding during collapse of their cores; mergers of neutron stars (NSs) probably represent the most popular progenitor model of short GRBs at the present time \cite{PiranReview2005,MeszarosReview2006}. 

GRBs are likely associated with a catastrophic energy release in stellar mass objects. The sudden emission of a large amount of energy in a compact volume (of the order of tens of kilometers cubed), leads to the formation of a relativistic ``fireball'' of $e^{\pm}$ pairs, $\gamma$-rays and baryons expanding in the form of a jet, while part of the gravitational energy liberated in the event is also converted into gravitational waves (GWs) \cite{PiranReview2005,MeszarosReview2006}. In the standard fireball model, the GRB electromagnetic emission is thought to be the result of kinetic energy dissipation within the relativistic flow, taking place at distances greater than $\sim10^{13}$~cm from the source \cite{PiranReview2005,MeszarosReview2006}. The electromagnetic signal can give indirect but important information on the progenitor's nature (e.g. its properties can constrain the structure and density of the circumburst medium, and it allows the identification of host galaxies). However, to reach a clearer understanding of the phenomenon, one should search for a direct signature of the progenitor's identity, which may be observed through the gravitational window. The energy that is expected to be radiated in GWs during the catastrophic event leading to a GRB would, in fact, be produced in the immediate neighborhood of the source. Thus, the observed GW signal would carry direct information on the properties of the progenitor.

In this paper, we present an analysis of the Virgo data \cite{VirgoWeb} simultaneous with the long GRB 050915a \cite{GRB050915aBAT}, with the goal to constrain the amplitude of a possible short burst of GWs associated with this GRB. At the time of GRB 050915a, Virgo was engaged in a 5-day long data run, named C7. Virgo's sensitivity during C7 exceeded that of all its previous runs. The lowest strain noise was $\sim 6\times10^{-22}$~Hz$^{-1/2}$ around $\sim 300$~Hz. This is the first time a study of this kind is being performed on Virgo data, so the work presented here aims also to define a procedure of analysis for GRB searches with Virgo. In the very near future, these kinds of studies will take advantage of the joint collaboration with LIGO, and the existence of an established procedure is fundamental. 

The sensitivity of Virgo during C7 was comparable to that of LIGO at the time when a coincidence search with GRB~030329 was performed \cite{Abbott030329}, and the upper-limits that we set for GRB 050915a are of the same order of magnitude, i.e. ${\cal O}\left(10^{-20}~{\rm Hz}^{-1/2}\right)$. The LIGO results on GRB~030329 thus represent a natural comparison for our analysis, even if procedures followed in this present study were developed for a single-detector search, while those of the LIGO study were for a double detector search.

In what follows, we first present an overview of the Virgo detector and its status during C7 (section \ref{Virgo}). Then, we recall the scenarios for GRB progenitors and their associated GW emission (section \ref{produzione di GW}), and we describe the main properties of GRB 050915a (section \ref{GRB050915a}). Furthermore, we present the analysis of Virgo data in coincidence with GRB 050915a (section \ref{pipeline}), and finally discuss our results (section \ref{discussione}).

\section{The Virgo detector}
\label{Virgo}
The Virgo gravitational wave detector, jointly funded by INFN (Italy) and CNRS (France), is located near Pisa, at the European Gravitational Observatory (EGO). It is a power recycled Michelson interferometer with $l=3$~km long arms, each containing a Fabry-Perot cavity (see e.g. \cite{GWDAWVirgoStatus} for a recent review of Virgo's status).

GW interferometric detectors like Virgo \cite{Rowan} have different types of source targets for their searches. These can be usefully divided in four main classes: stochastic waves, bursts, coalescing binaries and periodic waves. Sources that contribute to the first class are e.g. binary stars and primordial GWs (e.g. \cite{Maggiore,Weinberg,Montani}). Sources expected to produce GW signals of the other three classes are e.g. compact binaries and their coalescence (e.g. \cite{Blanchet} and references therein), rotating NSs with a non axis-symmetric mass distribution along the rotation axis or NSs instabilities (e.g. \cite{Ferrari2004a,Ferrari2004b,Manca}), collapse of massive stars and supernova (SN) explosions (e.g. \cite{FryerReview} and references therein). 
GRBs, short and long, are thought to be linked respectively to the coalescence of compact binaries and collapse of massive stars \cite{MeszarosReview2006}, and this has motivated searches for GWs signals in association with these sources [9, 21-37]\nocite{Abbott030329,ref1,ref2,ref3,ref4,ref5,ref6,ref7,ref8,ref9,ref10,ref11,ref12,ref13,ref14,ref15,ref16,ref17}.

\begin{figure}
\begin{center}
		\includegraphics[width=6cm,angle=90]{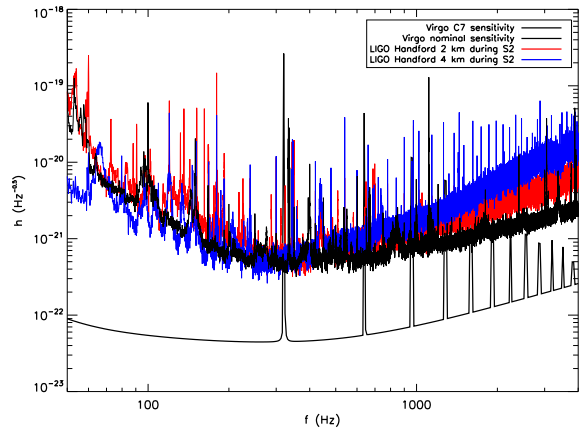}
		\caption[Virgo sensitivity during C7 run]{The Virgo sensitivity during C7 run and the Virgo nominal sensitivity are plotted in black. Typical LIGO Hanford sensitivities during the S2 are shown in red ($2$~km) and blue ($4$~km) \cite{LIGOWeb}.}
	\label{Fig1}
	\end{center}
\end{figure}

By the beginning of September 2005, Virgo was engaged in a commissioning run named C7, with the aim to test the gain in sensitivity after several improvements were performed on the detector (see \cite{VirgoStatus} for a review of Virgo status during C7). The run lasted 5 days, with a duty cycle of $\sim65\%$, and a mean sensitivity such that an optimally oriented 1.4 - 1.4 M$_{\odot}$ NS - NS binary coalescence, at a distance  of $\sim 1$~Mpc, would have been detected with a signal-to-noise ratio of 8. The noise spectrum of the detector during C7 may be roughly divided into two intervals. Below 300 Hz, the main contributions were from the noise associated with the control systems of the arms optical axes and of the angular degrees of freedom of the mirrors. Above 300 Hz, the dominant contribution was from shot noise \cite{VirgoNoise}. Figure \ref{Fig1} shows a comparison of the Virgo sensitivity curve during C7, and the LIGO sensitivity during its second science run (S2), when a search for a burst of GWs in coincidence with GRB~030329 was performed \cite{Abbott030329}. 

The Virgo data acquisition system acquires the interferometer signals, sampled at 20 kHz, and a large number of auxiliary signals pertaining to components of the interferometer or of its physical environment. This information is used to assess the state of various interferometer sub-systems (e.g. the mirror suspension status) and, in general, the quality of the collected data. A set of flags summarizes the data quality and, in particular, the ``Science mode'' flag indicates that the interferometer is in a stable locked configuration, and the modulation lines used for calibration purposes are properly set. The main physical information for the detection of GWs is extracted by reconstructing the interferometer strain. This fundamental step consists of the extraction of the arms' length relative difference $\delta l/l$, i.e. the amplitude of the GW signal, from the output dark fringe signal \cite{TesiBeauville}. The last is corrected for the effects of controls by subtracting the contribution of the pendulum motion of the suspended mirrors, caused by injected control signals on the reference mass coil drivers. Since the reconstruction procedure performs a subtraction of the control signals, it removes part of the injected control noise and tends to cancel the calibration lines, like any signal injected in the mirror control. In practice, the goodness of the calibration lines subtraction is used to monitor the quality of the reconstruction. During the C7 run, the error in the $h$-reconstructed data ($h(t)=\delta l/l$) was estimated to be around $\sim +20\%-40\%$. Hereafter, this is assumed as a systematic error in our analysis.

\section{GRB progenitor models and the expected GW signal}
\label{produzione di GW}

The actual favored scenario for long GRB progenitors is the so-called ``collapsar'' model, which invokes the collapse of a massive star down to a black hole (BH) with formation of an accretion disk, in a peculiar type of SN-like explosion (see e.g. \cite{Woosley1993,Paczynski1998,Fryer1999}). On the other hand, the favored scenario for short GRB progenitors is the compact binary (NS - NS or BH - NS) merger. This process is believed to occur extremely quickly and be completely over within a few seconds, naturally accounting for the short nature of these bursts. Unlike with long bursts, there is no conventional star to explode and therefore no SN is expected. To produce a GRB, both long or short, it is required that the progenitor stellar system ends as a rotating BH and a massive disk of matter around it, whose accretion powers the GRB ultra-relativistic fireball in the form of a jet, along the rotational axis of the system \cite{PiranReview2005}. Due to the relativistic beaming effect, only observers located within the jet opening angle are able to observe the emission from the jet. In the standard assumption (see e.g. \cite{Frail2001}), the jets are uniform, with sharp cut-offs at the edges, and the line of sight cuts right across the jet axis, i.e. the Earth is near the center of the $\gamma$-ray beam. 

GWs are expected to be emitted in association with both long and short GRBs \cite{MeszarosReview1999,Kobayashi2003}. Coalescing binaries, thought to be associated with short bursts, are one of the most promising GW sources for interferometric detectors like Virgo. For such systems, a chirp signal should be emitted in GWs during the in-spiral, followed by a burst-type signal associated with the merger and subsequently a signal from the ring-down phase of the newly formed BH. The last, initially deformed, is expected to radiate GWs until reaching a Kerr geometry \cite{Kobayashi2003}. In the collapsar scenario, relevant for long GRBs, the high rotation required to form the centrifugally supported disk that powers the GRB, should produce GWs emission via bar or fragmentation instabilities that might develop in the collapsing core and/or in the disk \cite{Kobayashi2003}. Moreover, asymmetrically in-falling matter is expected to perturb the final BH geometry, leading to a ring-down phase \cite{Kobayashi2003}. While an axis-symmetric rotating collapse and core bounce would give no contribution to GW emission along the GRB axis, bar and fragmentation instabilities are all dominated by modes with spherical harmonic indices $l=m=2$ \cite{Kobpolarization}, implying that GRB progenitors would emit more strongly along the GRB axis than in the equatorial plane (i.e. the orbital plane of the disk fragments or of the bar). The same holds for the ringing BH \cite{Kobpolarization}. Thus, in the standard scenario having the Earth near the center of the $\gamma$-ray beam, the detector is located in the maximum of the emission pattern for GWs dominated by spherical harmonic indices $l=m=2$.

Simulations of GW emission in the process of core-collapse in massive stars (see e.g. \cite{FryerReview} and references therein), has advanced much recently. However, their applicability to GRBs is not clear, since in these simulations
the final outcome of the collapse are generally NSs. Moreover, those simulations typically do not arrive at
the point when non-axis-symmetric instabilities are developed. Recently, \cite{Dimmelmeier} have further extended the post-bounce evolution of some pre-supernova models characterized by a relatively quick spinning stellar iron core, finding that these models do indeed develop a considerable non-axis-symmetry after the bounce, with associated GW emission in agreement with the expectations from a spinning-bar model \cite{Dimmelmeier}. The study of GW emission by gravitational collapse of uniformly rotating NSs to BHs, is also advancing (see e.g. \cite{Baiotti}). However, due to the high degree of axis-symmetry of these simulations, their outcomes do not represent the best scenario for GRB progenitors. In this article, we assume a model based on a best case scenario, and work under the hypothesis that bar and fragmentation instabilities do indeed develop, and that the Earth (i.e. the Virgo detector) is aligned with the center of GRB~050915a $\gamma$-ray beam. 

A final important aspect is the expected delay between the electromagnetic trigger and the associated GW signal. Typically, a GW signal is searched for within a window of $180$~s around the GRB trigger. For a GW burst associated with the formation of the GRB central engine, i.e. with the launch of the jet that subsequently powers the GRB, the electromagnetic trigger should follow the GW trigger. As described in \cite{mes04}, in the case of a collapsar (relevant for our analysis), the time delay between the two triggers is dominated by the time necessary for the fireball to push through the stellar envelop of the progenitor, which can be of the order of $10-100$~s. Thus, a period of $120$~s before the trigger time is typically selected to over-cover these predictions. Moreover, given that some models predict a GW signal contemporaneous with the GRB emission (i.e. extending from the time of onset of the GRB to its end, see \cite{Putten}), the $60$~s after the electromagnetic trigger are also included in our search.

\section{GRB 050915a}
\label{GRB050915a}

On the 15$^{\rm th}$ of September 2005, at T$=11:22:42$~UT, the ``Burst Alert Telescope'' (BAT) on-board \textit{Swift} \cite{Gehrels2004} triggered and located GRB~050915a \cite{GRB050915aBAT}.  The BAT on-board calculated position was RA=$05$h~$26$m~$51$s, Dec=$-28$d~$01'~48''$ (J2000), with an uncertainty of $3$~arcmin. The BAT measured a T$_{90}$ \footnote{The $T_{90}$ duration is defined as the time necessary to collect from $5\%$ to $95\%$ of the total counts in the specified energy band.} duration of $53\pm3$~s in the $15-350$~keV energy band \cite{GRB050915aBATref}, thus GRB~050915a was classified as a long-type GRB. The ``X-Ray Telescope'' (XRT) began observing the BAT position at $11:24:09$~UT ($\sim 87$~s after the trigger, \cite{GRB050915aXRT}). The new refined position was RA=$05$h$~26$m~$44.6$s, DEC=$-28$d~$01$m~$01.0$s \cite{GRB050915aXRT}. Finally, the \textit{Swift} ``Ultra-Violet and Optical Telescope'' (UVOT) began observing the field of GRB050915 $\sim85$~s after the BAT trigger \cite{GCN3986}, and the optical and IR follow-up of this burst was performed by different telescopes \cite{GCN3978,GCN3980,GCN3981,GCN3985,GCN3984,GCN3990}. In the radio band, VLA observations on September $18.58$~UT revealed no radio source in the error circle \cite{GCN4001}. Recently, evidence has been found for a possible distant and/or faint galaxy \cite{Jakobsson,astro-ph/0703388}, however the redshift of this burst still remains unknown.

\section{Coincidence analysis for a burst of GWs}
\label{pipeline}
Given the lack of accurate predictions on the expected GW waveforms that might be associated with long GRB progenitors, we have chosen in this search to look for a GW burst-type signal associated with GRB~050915a in a model independent way. Our analysis aims to:
\begin{itemize}
\item set-up a procedure for GRB triggered searches with Virgo, that goes well beyond the single case of GRB~050915a, but has a much more general interest, both in the context of GWs triggered searches and GRB research, and also in view of the future joint collaboration with LIGO;
\item constrain the amplitude of the associated GW emission to quantify Virgo's capability to provide information pertaining to models for GW production by GRBs, and also to define possible margins of improvement in view of the expected enhancement in sensitivity;
\item define an algorithm that will find application also for the categories of short GRBs (merger and ring-down phase, see section \ref{produzione di GW}), which are expected to be nearer and thus more promising sources of GWs than long bursts.
\end{itemize}

\subsection{Wavelet Analysis}
For our analysis, we relied on a new wavelet-based transient detection tool, the Wavelet Detection Filter (WDF). Wavelets were introduced in the
80's as a mathematical tool to represent data both in time and
frequency \cite{daubechies92:_ten_lectur_wavel}. The wavelet transform is defined as the correlation of the
data $x(t)$ against the wavelets $\psi_{a,b}$,
\begin{equation}
W_x(a,b)=\int_{-\infty}^{+\infty} x(t) \psi_{a,b}(t) dt.
\end{equation}
The wavelet family is obtained by translations and dilations
of a reference waveform $\psi$
\begin{equation}
\psi_{a,b}(t)=\frac{1}{\sqrt{a}} \psi\left(\frac{t-b}{a}\right).
\end{equation}
The wavelet transform gives a representation of the signal in terms
of the scale $a$ (associated to frequency)  and time $b$. The reference wavelet $\psi(t')$ is chosen to be a zero-mean function of unit energy, well-localized both in time (around $t'=0$) and frequency. Consequently, this is a short-duration waveform with few cycles. Wavelet-based representations are well-suited for burst-like signals because of similarities between those signals and the analyzing wavelet $\psi$.

The wavelet family is redundant but when the sampling of the scale $a$
and time $b$ axes is dyadic i.e., when $(a_j, b_k)=(2^j, k 2^j)$
for $j\geq 0$ and $k$ integers, it forms an orthonormal basis, provided
some geometrical constraints on the choice of $\psi$ are set (we will not
detail them here, the reader is referred to 
\cite{daubechies92:_ten_lectur_wavel}). The wavelet transform produced
in this way is referred to as a \textit{discrete wavelet transform}, and we
use it to analyze the data. Since we are dealing with signals sampled at a rate of $f_s=20$ kHz, and we
consider data blocks of $N$ samples, the dyadic sampling is limited to
the range $j=0,\ldots,\log_2 N$ and $b_k/a_j=k/f_s$ with
$k=0,\ldots,N-1$. 

\subsection{Best matching wavelets and thresholding}

The wavelet transform can be viewed as a bank of matched filters. We
select the best matching wavelets (largest correlation coefficients) by thresholding. Let us define the soft-thresholding operator
$T(w;\eta)= sign(w) (|w|-\eta)$ if $|w|>\eta$ and 0 otherwise,
and the thresholded coefficients of the discrete wavelet transform
$w_{j,k}=T(W_x(a_j,b_k);\eta)$.

We use the Signal to Noise Ratio ($S_{w}$) as a statistic 
\begin{equation}
S_{w} = \sqrt{\frac{\sum_{j,k} w^2_{j,k}}{\sigma^2_{n}}}  
\end{equation}
to discriminate the presence or absence of a
burst-like signal in the data, with $\sigma^2_{n}$ being an estimate of the variance of the noise. The threshold choice is $\eta= \sqrt{2 \log N}\sigma_{n}$. This choice is linked to a general result by Donoho \&
Johnstone \cite{Dohono} concerning non-parametric denoising. Let our data
$x(t)=s(t)+n(t)$ be the sum of a signal $s(t)$ and Gaussian white
noise $n(t)$. It can be shown that the signal estimate obtained as 
\begin{equation}
\hat{s}(t)=\sum_{j,k}
w_{j,k} \psi_{a_j,b_k}(t)
\end{equation}
i.e. by inverting the thresholded wavelet transform, minimizes the mean-square error
over a broad class of signals \cite{Mallat}.

\subsection{Pipeline: preprocessing, analysis and trigger selection}

\begin{figure}
\begin{center}
		\includegraphics[width=6cm, angle=90]{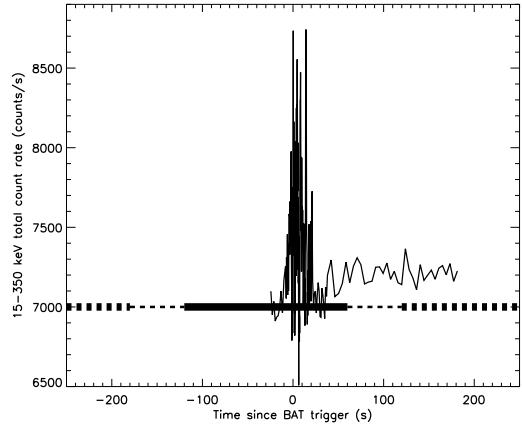}
		\caption{BAT (15-350 keV) light curve of GRB 050915a in total count rate (counts/s, thin solid line). Data of the BAT light curve for GRB~050915a has been downloaded from the online archive \cite{BATonline}. The thick solid line defines the time length and position of the on-source region, the thin-dashed segments mark the $60$ s of data before the start and after the end of the signal region which are excluded from the analysis (so to separate the background from the signal); the thick-dashed segments mark the portions of the off-source region around the GRB trigger time. The whole off-source region considered in our analysis is much longer, extending on the left of the plot up to $-9973$~s, and on the right of the plot up to $6863$~s.}
	\label{GWDAWfig}
	\end{center}
\end{figure}

We preprocess the data by applying a time-domain whitening procedure \cite{NotaElena1} using an estimate of the noise spectral density given by an autoregressive (AR) fit over the first 1000 s of the data set. We divide the whitened time-series into overlapping blocks of duration $\Delta t_W=12.8$~ms, which also sets the time resolution of the search. The epochs of two successive blocks differ by
$\Delta t_s=0.6$~ms. 

We already mentioned that the wavelet basis may be interpreted as a template grid. 
It is well-known that redundant grids are better suited for burst detection because 
they increase the chance of a good match between the signal and one of the templates. 
Instead, the wavelet bases are sparse by construction. To compensate for this sparsity, 
we insert redundancy by combining the results from several wavelet decompositions. 
We empirically select 18 different wavelets (including Daubechies wavelets from 4 to 
20 \cite{daubechies92:_ten_lectur_wavel}, the Haar wavelet and the windowed Discrete Cosine \cite{Mallat}). For each data block, we compute these 18 wavelet transforms. Thus, we obtain 18 $S_{w}$ estimates and select the largest one ($SNR$). Furthermore, we generate a shorter trigger list by selecting only $SNR$ values larger than 4.

Typically, the energy content of a burst signal will be tracked by the WDF in the form 
of a ``cluster'' of successive triggers, each containing a different fraction of its energy. To assign to each candidate event a unique time and $SNR$ value, such clustering should be properly organized. As such, we cluster all triggers having a time difference less than $10$~ms ($\sim \Delta t_W$). For each cluster, a candidate event is defined having the time and $SNR$ of the trigger at which the maximum $SNR$ of the cluster is reached, i.e. where the signal leaves most of its energy.

\subsection{The analysis method}
For our analysis, we rely on a single stretch of data during which the interferometer was maintained in the same configuration and the $h$-reconstruction processes was recognized as good. Such a stretch is between the GPS times $810808602$~s and $810825638$~s, for a total of $16836$~s, containing the GRB trigger time (i.e. GPS time $\sim 810818575$~s). We define as the on-source region a data segment $180$~s long, $120$~s before the GRB trigger time and $60$~s after (see the thick-solid line in Fig. \ref{GWDAWfig}). This is the time window where we searched for a coincidence with the GRB trigger. The rest of the data in the stretch, with the exception $60$~s before the start and after the end of the signal region (see the thin-dashed in Fig. \ref{GWDAWfig}), belong to what we define as the off-source region (see the thick-dashed lines in Fig. \ref{GWDAWfig}). The analysis of the off-source region is used to assess the data quality and to study the statistical properties of the background. 
As explained in section \ref{produzione di GW}, the on-source region has been chosen to start $120$~s before the trigger time, so as to over-cover most of astrophysical predictions regarding the expected delay between the GRB and the associated burst-type GW signal. Moreover, in view of models that predict a GW signal contemporaneous with the GRB emission and considering that GRB~050915a had a T$_{90}$ duration of $53\pm3$~s (see section \ref{GRB050915a}), we have chosen our signal region to end $60$~s after the trigger. The same choice for the time-length of the on-source region was also implemented in \cite{Abbott030329}, for the case of the long GRB~030329.

Our pipeline is calibrated by adding simulated signals of various amplitudes and waveforms to data in the off-source region. The simulated signals were produced using the Virgo SIESTA simulation code \cite{siesta}. The resulting data stream is processed in the same way as for the off- and on-source regions. Using simulated signals we evaluate the detection efficiency as a function of the simulated signal strength, which we quantify in terms of root-sum square amplitude of the incoherent sum of the contributions from the ``plus'' and ``cross'' polarizations:
\begin{equation}
h_{rss}=\sqrt{\int^{+\infty}_{-\infty}\left(h^{2}_+(t)+h^{2}_{\times}(t)\right)dt}.
\label{hrss}
\end{equation}
This allows us to make physical interpretations with the results.

The calibration procedure is based on simulations of plausible but quite general burst-type waveforms, with different amplitudes, characteristic frequencies and durations, chosen on the basis of the considerations explained in section \ref{GW emission}. The times at which those signals are added to the off-source data are randomly determined by following a Poisson distribution with a mean rate of $0.1$~Hz. 

Knowledge of the source position is also used when adding the simulated signals to off-source data, by considering the antenna response at the GRB position and time. A GW arriving at the interferometer from the GRB direction can  be described as a superposition of two polarizations amplitudes $h_{+}$ and $h_{\times}$. The response of the interferometer to such a wave is given by \cite{Thorne}:
\begin{equation}
	\delta l/l=h(t)=F_{+}h_+(t)+F_{\times}h_{\times}(t)
\end{equation}
where $F_{+}$ and $F_{\times}$ are expressed as functions of the source position and of the wave polarization angle $\psi$ \cite{FinnChernoffphysrevD1993_47_2198}. The antenna patter functions $F_{+}$ and $F_{\times}$ can be written as:
\begin{eqnarray}
\label{risposta+}F_+=F^{0}_{+}\cos(2\psi)-F^{0}_{\times}\sin(2\psi)\\
F_{\times}=F^{0}_{+}\sin(2\psi)+F^{0}_{\times}\cos(2\psi)
\label{rispostax}
\end{eqnarray}
where $F_{+}(\psi=0)=F^{0}_{+}$ and $F_{\times}(\psi=0)=F^{0}_{\times}$. In the case of GRB~050915a, we have $F^{0}_{+}\sim 0.32$ and $F^{0}_{\times}\sim 0.21$.

\subsection{Choice of plausible waveforms}
\label{GW emission}
\subsubsection{Gaussian waveforms}
To calibrate our pipeline, considering the great uncertainties in the waveforms associated with long GRB progenitors, we added different burst-type signals to the off-source data. Our simplest choice was for Gaussian signals, having the following form:
\begin{equation}
	h(t)=h_{0}\exp[-(t-t_0)^{2}/2\sigma^{2}] F^0_+
	\label{gaussiane}
\end{equation}
where $t_0$ is the time at which the signal is added to the off-source data stream, and $\sigma$ values of $0.5$~ms, $1$~ms and $1.5$~ms were considered. These were broad-band, linearly polarized waveforms along the $+$ direction, with the unknown polarization angle $\psi$ set to zero. Note that for a given value of $\sigma$ and $h_0$, a different choice of $\psi$ rescales the waveform amplitude arriving at the detector by a factor of $\cos(2\psi)$, while leaving unchanged its shape over the detector bandwidth. Thus, the efficiency curves for the general $\psi$ case can be estimated from the $\psi=0$ ones here presented, by rescaling the $h_{rss}$ corresponding to a given detection efficiency for a factor of $\frac{1}{\cos(2\psi)}$.

\subsubsection{Sine-Gaussian waveforms}
To mimic GW emission by GRB progenitors during the phase of collapse, fragmentation or bar instabilities, we considered sine-Gaussian waveforms. Taking a best case model scenario of GW emission from a triaxial ellipsoid rotating about the same axis as the GRB (i.e., the direction to the Earth, see Eqs. (\ref{hxfinale}) and (\ref{h+finale}) in the appendix \ref{appendice}), and using a Gaussian amplitude as the simplest way to mimic the impulsive character of a GW burst, we consider signals having in the wave frame the following form:
\begin{eqnarray}
h_+=h_0\exp[-(t-t_0)^{2}/2\sigma^{2}]\cos(2\pi f_0 (t-t_0))\label{piu}\\
h_{\times}=h_0\exp[-(t-t_0)^{2}/2\sigma^{2}]\sin(2\pi f_0 (t-t_0))\label{per}
\end{eqnarray}
for an unknown value of polarization angle $\psi$. The detector response to such types of signals is then computed using Eqs. (\ref{risposta+})-(\ref{rispostax}). After some algebra, the resulting $h(t)$ can be written as:
\begin{eqnarray}
\nonumber h(t)=h_{0}\exp[-(t-t_0)^{2}/2\sigma^{2}][F^{0}_+\cos(2\pi f_0 (t-t_0) -2\psi)+\\+F^{0}_{\times}\sin(2\pi f_0 (t-t_0) - 2\psi)]
\label{indip}
\end{eqnarray}
In our analysis we set $\psi=0$ and we span the frequency range $f_0\sim 200-1500$~Hz, as suggested by the predictions for GW emission from GRB progenitors, when fragmentation or bar instabilities are developed (see dash-dotted lines in Figs. 3-5 of \cite{Kobayashi2003}). For each $f_0$, we consider two values of $Q$, i.e. $Q=5$ and $Q=15$.

It is worth noting that, for signals of the form (\ref{piu})-(\ref{per}) with $Q=2\pi f_0 \sigma\gtrsim 3$ (i.e. for relatively narrow-band signals), one has:
\begin{equation}
	h_{rss}\simeq \sqrt{h_0^2\frac{Q}{2\sqrt{\pi} f_0}}
\end{equation}
and:
\begin{equation}
\sqrt{\int_{-\infty}^{+\infty}h^2(t)dt}\simeq h_{rss}\sqrt{\frac{(F^0_+)^2+(F^0_x)^2}{2}}
\end{equation}
where $h(t)$ is given by Eq. (\ref{indip}). In this approximation, if the detector noise is roughly constant within the relatively narrow signal bandwidth, the detected $SNR$ is proportional to the above integral. Thus, the detection efficiency as a function of the $h_{rss}$ is expected to be independent on the choice of $\psi$.

\subsubsection{Damped sinusoid waveforms}

Consider now the phase of BH ringing. A Kerr BH distortion can be decomposed into spheroidal modes with spherical-harmonic-like indices $l$ and $m$ (see e.g. \cite{fryerholzhughes}). The quadrupole modes $(l=2)$ presumably dominate \cite{fryerholzhughes}, while the paramount $m$-value depends upon the matter flow. In particular, the $m=\pm2$ modes are bar-like, co-rotating $(m=+2)$ and counter-rotating $(m=-2)$ with the BH spin, and the $l=m=2$ mode is expected to be the most slowly damped one. 
As underlined in \cite{Berti}, numerical simulations of a variety of dynamical processes involving BHs show that, at intermediate times, the response of a BH is indeed well described by a linear superposition of damped exponentials. Generally speaking, the polarization of the ring-down waveform will depend on the physical process generating the distortion of the BH (see e.g. \cite{FerrariBH}). Since the $l=m=2$ mode may be preferentially excited in the presence of binary masses or fragmentation of a massive disk, it is commonly assumed that the distribution of the strain between polarizations $h_{+}$, $h_{\times}$ for this mode mimics that of the in-spiral phase \cite{Kobpolarization,rhook2005,Berti},
\begin{equation}
h_{+}=h_{0}\frac{1}{2}(1+\cos^{2}\theta_0)\exp(-t/\tau)\cos(2\pi f_0 t+\xi)\Theta(t),
\end{equation}
\begin{equation} 
h_{\times}=h_{0}\cos\theta_0\exp(-t/\tau)\sin(2 \pi f_0 t+\xi)\Theta(t),
\end{equation} 
where $\xi$ is an arbitrary phase, $\theta_0$ is the inclination of the angular momentum axis with respect to the source direction in the sky, and $\Theta(t)$ is the normalized step function. Since we expect to observe the GRB on-axis, this polarization is also circular.

Thus, we added to the off-source region signals having the form 
\begin{eqnarray}
\nonumber h(t)=h_{0}\exp[-(t-t_0)/\tau]\times \\\nonumber\times[F^{0}_{+}\cos(2\pi f_0(t-t_0))\Theta\left(1-\frac{1}{4 f_0(t-t_0)}\right)\Theta(t-t_0)+\\+F^{0}_{\times}\sin(2\pi f_0(t-t_0))\Theta(t-t_0)],
\label{imp1}
\end{eqnarray} 
where again $\Theta(t-t_0)$ and $\Theta\left(1-\frac{1}{4 f_{0}(t-t_0)}\right)$ are normalized step functions \footnote{The reason for multiplying $\cos(2\pi f_{0}(t-t_0))$ by $\Theta\left(1-\frac{1}{4 f_{0}(t-t_0)}\right)$ is to avoid a discontinuity at the beginning of the waveform, which would result in an infinite energy, even though $h_{rss}$ would remain finite.}.

The characteristic frequency of the $l=m=2$ quasi-normal mode of a Kerr BH is estimated as \cite{Kobayashi2003}
\begin{equation}
	f_{0}=32~{\rm kHz}~(1-0.63(1-a)^{3/10})\left(\frac{M}{M_{\odot}}\right)^{-1},
	\label{freqringdown}
\end{equation}
where $a$ is the dimensionless spin parameter of the Kerr BH, while the damping time can be estimated as
\begin{equation}
	\tau\sim (\Delta f)^{-1}=\pi f_{0}/Q(a),
	\label{eqtau}
\end{equation}
where $Q(a)=2(1-a)^{-9/20}$ \cite{Kobayashi2003,Echeverria1989}. For GRB progenitors it is typically assumed $a=0.98$, since the BH is supposed to have spun up to near maximal rotation by a massive accretion disk \cite{Kobayashi2003}. In the collapsar model, $M$ is expected to be of the order of $\sim 1$~M$_{\odot}$, which implies $f_{0}\sim 10$ kHz \cite{Kobayashi2003}, so a detection would be difficult. 

However, the collapsar model for long GRBs is one of a larger class of proposed progenitor models, all leading to a final BH plus accretion disk system. Thus, the process of GW emission can always be described in a way similar to the collapsar case \cite{Kobayashi2003}: a high rotation rate producing bar or fragmentation instabilities in the disk, followed by a BH initially deformed encountering a ring-down phase. Among these variants of the collapsar model, the BH-white dwarf scenario may be characterized by higher BH masses ($M \sim 10$~M$_{\odot}$) and lower $f_0$ values, down to $\sim 1$~kHz (see Fig. 3 of \cite{Kobayashi2003}). According to Eq. (\ref{eqtau}), a $f_0$ around $1$~kHz would imply a damping time of $\sim 0.3$~ms. Thus, we choose to span a frequency range between $\sim 800$~Hz and $3$~kHz, for $\tau$ values of $0.3$~ms, $1$~ms and $1.5$~ms.

\section{Results and Discussion}
\label{discussione}

\subsection{Data Quality}
\label{nuova_sec_DQ}
We applied the WDF to the data of the off-source region and we derived 
the distribution of the trigger strength. In the off-source region, we 
selected $\sim 2.1 \times 10^{3}$ triggers crossing the threshold $SNR = 
4$. We processed the resulting list so as to eliminate triggers related to instrumental 
artifacts. The trigger rejection operates in two steps \cite{allsky}, namely a 
data preselection followed by a glitch removal procedure, that we can summarize as follows.

\begin{figure}
\begin{center}
		\includegraphics[width=6cm,angle=90]{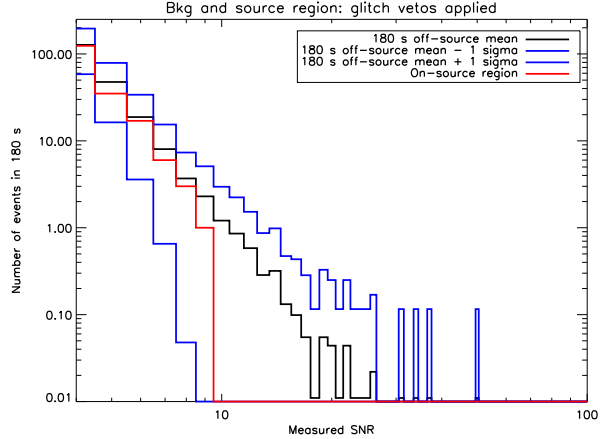}
		\caption[]{The mean $SNR$ distribution found in the off-source region (black) and its $\pm1\sigma$ interval (blue) is compared with the on-source $SNR$ distribution (red). As is evident, the on-source distribution is within the $1\sigma$ interval around the mean off-source one, confirming that the on-source and off-source distributions are statistically compatible.}
	\label{meanhisto}
	\end{center}
\end{figure}

First, triggers falling into periods where known instrumental problems occurred (e.g. saturation of the control loop electronics, problems in the $h$-reconstruction process) or when aircraft (known to produce high seismic/acoustic noise which couples into the interferometer) fly over the instrument, are discarded. This preselection, while leaving the on-source region untouched, cuts out from the off-source segment (16500 s) about 14 s associated with high acoustic noise, occurring about 1945 s after the start of the off-source stretch. Furthermore, a glitch removal procedure was applied. An extensive study has been performed to establish the correlation of triggers produced by the burst search pipelines, and environmental or instrumental glitches occurring during C7 (see \cite{allsky} for a detailed description). This study provided the definition of a series of veto criteria, based on information given by the auxiliary channels, and introduced a ``dead time'' of $\sim 6.3\%$ \cite{allsky} on the complete data set of the C7 run.  The application of these criteria in our analysis leads to a dead time of $\sim 582$ s in the off-source region (i.e. $\sim 3.5 \%$ of its duration), and of $\sim 1.3$ s in the on-source region (i.e. $\sim 0.75 \%$ of its duration). The loudest on-source event remains unaffected by the veto procedure. 

It is important to note that we have {\it not} applied {\it all} the vetoes designed for the C7 data. In fact, the majority of spurious burst triggers has been related to the so-called ``Burst of Burst'' (BoB) \cite{allsky}. BoBs originate from a misalignment of the interferometer mirrors, which increases the coupling of the laser frequency noise into the interferometer itself and causes a noise increase lasting up to a few seconds. Procedures have been defined to veto the BoBs. However, as estimated from the complete dataset of C7 run, these introduce a large dead  time ($\sim16\%$ of the run duration \cite{allsky}). Thus, since the BoB cuts can significantly affect the integrity of the on-source segment, we chose \textit{not} to apply them. It is worth noting that during a BoB, the interferometer is sensitive to GWs, therefore we can still observe a possible GW counterpart of a GRB, albeit with a lower $SNR$.

\subsection{Statistical Analysis}
\label{DQ}

\begin{figure}
\begin{center}
		\includegraphics[width=6cm,angle=90]{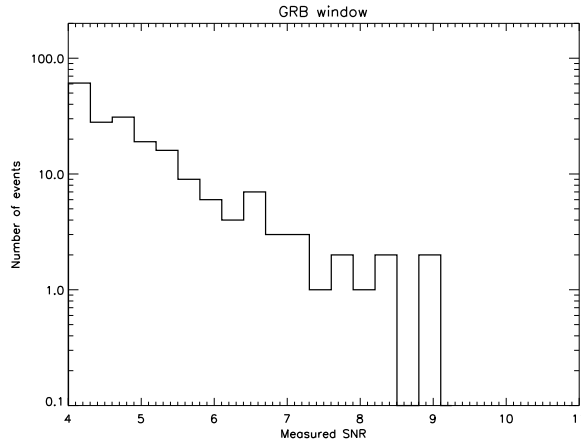}
		\caption[]{Distribution of event strengths in the on-source region (number of events vs detected $SNR$).}
	\label{istogrammaGRB}
	\end{center}
\end{figure}

We show in Fig. \ref{meanhisto} the $SNR$ distribution found in the off-source and on-source regions, where we have applied the data quality cuts described in section \ref{nuova_sec_DQ}. The $SNR$ distribution in the on-source region is confined below $SNR=9$ (see Fig. \ref{istogrammaGRB}). To test if the distribution of events observed on-source is compatible with being only noise, we performed different checks. First, starting from the beginning of our data stretch (i.e. GPS $810808602$~s), we sampled our background distribution by dividing the off-source region in $\sim 90$ successive windows $180$~s long. We find that the percentage of such windows having a loudest event with $SNR$ $> 9$ is of about $89\%$ (after applying in each of the windows the cuts described in the prevoious section). Second, we computed the mean $SNR$ distribution on the $90$ off-source windows, and derived for each $SNR$ bin the corresponding $\sigma$. As shown in Fig. \ref{meanhisto}, the on-source distribution is well within the $\pm 1\sigma$ interval around the mean off-source one, thus being compatible with noise. 

From these tests we conclude that the on-source events are consistent with noise and that no clear evidence is found for an exceptional event, with respect to the background statistics, that could possibly be associated with the GRB. Thus, we move to the definition of an upper-limit, by following the procedure described in \cite{loudestref}. To this end, we use simulated signals to determine the strain necessary to have $90\%$ frequentist probability for such signals showing up as events with $SNR>9$, i.e. with a $SNR$ above that of the loudest event observed in the on-source region. This means estimating the efficiency $\epsilon$ at which the instrument and filtering process can detect burst events with $SNR>9$. According to \cite{loudestref}, the simulated signals are added to the \textit{off-source data}, so as to evaluate $\epsilon$ with good statistics, thanks to the long duration of the chosen off-source stream. Before proceeding with the efficiency estimates, basic sanity checks aimed to guarantee the consistency of our approach were applied. These included a comparison of the off-source and on-source regions in the time-frequency domain, and a Kolmogorov-Smirnov test ($90\%$ confidence level, two-sided test) between the on- and off-source $SNR$ distributions.

\subsection{Detection efficiency and upper-limit strain}
\label{deteff}
The efficiency $\epsilon$ in detecting signals with $SNR>9$, is estimated by computing for each kind of chosen waveform (see section \ref{GW emission}), the percentage of simulated signals found by the pipeline, as a function of the injected strain amplitude $h_{rss}$. A simulated signal added to the noise at a given time $t_0$, is recognized by the pipeline at $t_{det}$ if $\left|t_{det}-t_0\right|\leq 20$~ms. This $\pm 20$~ms coincidence window takes into account the duration of the wavelet decomposition window ($12.8$~ms), allowing a partial overlap. Moreover, a coincidence window of $\pm 20$~ms contains the $\sim \pm2\sigma$ portion of the longest duration simulated signal (sine-Gaussian waveforms with $Q=15$ and frequency $203$~Hz, having $\sigma\sim 12$~ms).      

\begin{figure}
\begin{center}
		\includegraphics[width=6cm,angle=90]{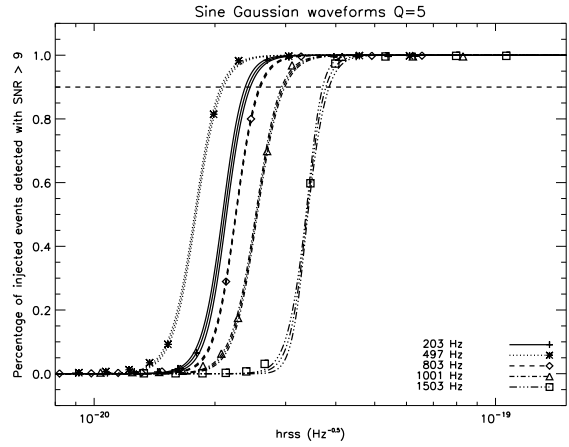}
		\caption[]{Efficiency in detecting sine-Gaussian signals with $Q=5$, for different characteristic frequencies $f_0$, as a function of the injected $h_{rss}$. Only events found with $SNR>9$ and within $\pm20$~ms the injection time are counted in the measured efficiency. The dashed line marks the $90\%$ efficiency level. For each waveform we plot the measured efficiency (points), the best fit sigmoid function and the other two curves which account for fitting errors (in most cases, these two curves appear overlapped on the best fit one).}
	\label{SGQ5eff}
	\end{center}
\end{figure}

\begin{figure}
\begin{center}
		\includegraphics[width=6cm,angle=90]{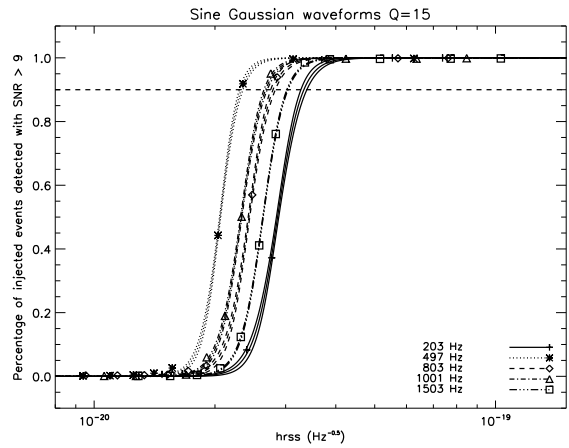}
		\caption[]{Similar to Fig. \ref{SGQ5eff}, efficiency curves for sine-Gaussian waveforms with $Q=15$ (see the caption of Fig. \ref{SGQ5eff} for further details).}
	\label{SGQ15eff}
	\end{center}
\end{figure}

In Fig. \ref{SGQ5eff} and Fig. \ref{SGQ15eff} we show the results obtained for sine-Gaussian waveforms with $Q=5$ and $Q=15$ respectively, at different frequencies. As one can see, the detection efficiency depends on the signal frequency. There are two main elements determining such dependence: (i) the detector noise level and (ii) the signal duration with respect to the window in which the wavelet decomposition is performed. Concerning point (i), consider two sine-Gaussian signals at different frequencies, with equal strain amplitude $h_{rss}$. Those will be detected at different $SNR$ values, since the detector noise level changes with frequency. Thus, the lowest is the detector noise around the signal characteristic frequency, the highest will be its detection efficiency. Concerning point (ii), sine-Gaussian waveforms with the same $Q$ but differing $f_0$ have different durations (i.e.  $5\sigma=5\frac{Q}{2\pi f_0}$). The filter capability in detecting a signal is maximized when the duration of the window in which the wavelet decomposition is performed, is comparable to the signal duration. When taking a shorter wavelet window, part of the signal power is lost. On the other hand, choosing a wavelet window much longer than the signal duration, implies that the wavelet decomposition is dominated by the background, resulting in a loss of efficiency. This is the reason why in our analysis, as a trade off, we set a wavelet decomposition window of $12.8$~ms, comparable to the $\sigma\sim 12$~ms of the longest injected event. 


\begin{figure}
\begin{center}
		\includegraphics[width=6cm,angle=90]{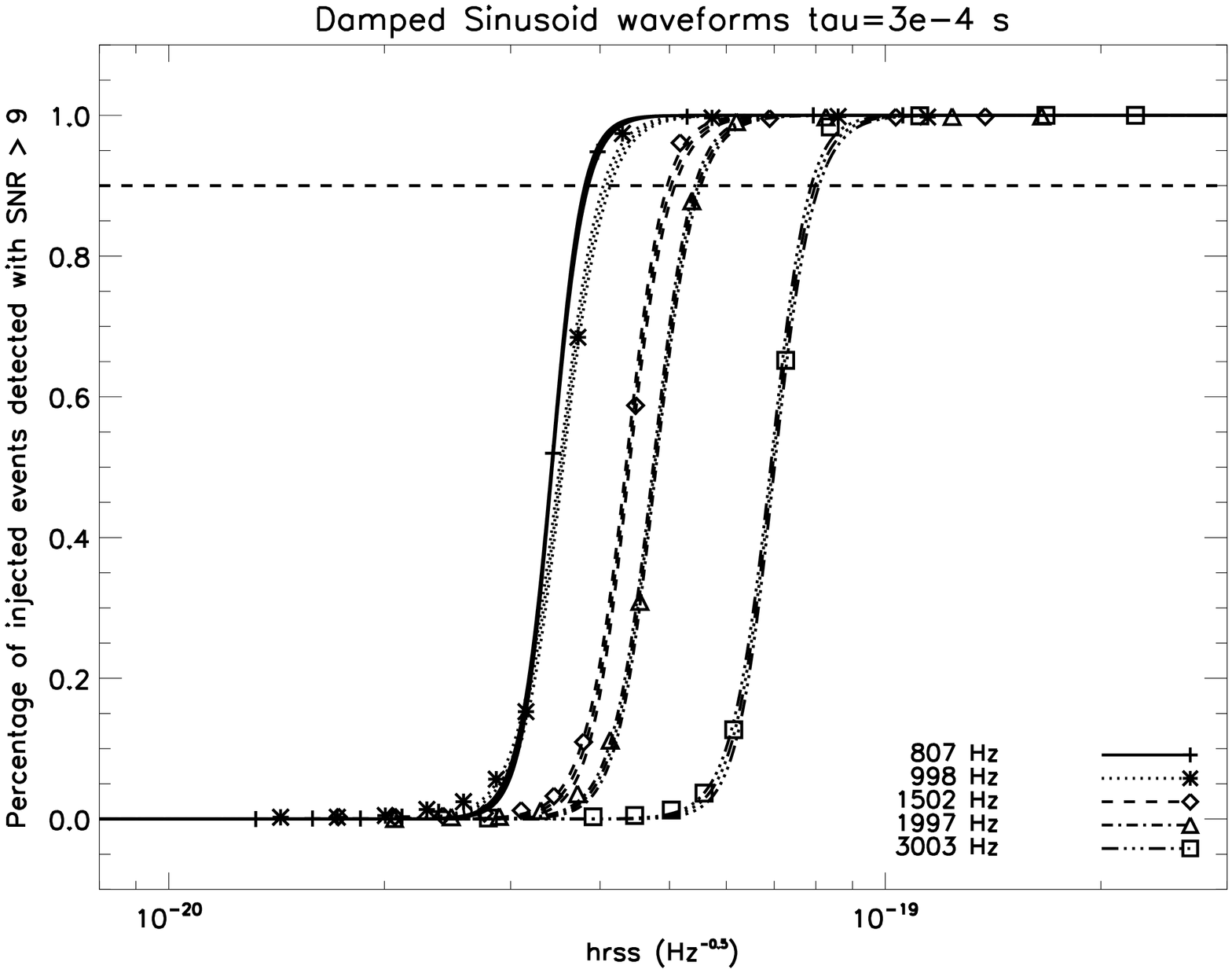}
		\caption[]{Efficiency curves for damped-sinusoid waveforms with damping time $\tau=0.3$~ms, for different characteristic frequencies $f_0$, as a function of the injected $h_{rss}$ (see the caption of Fig. \ref{SGQ5eff} for further details).}
	\label{DS3}
	\end{center}
\end{figure}

\begin{figure}
\begin{center}
		\includegraphics[width=6cm,angle=90]{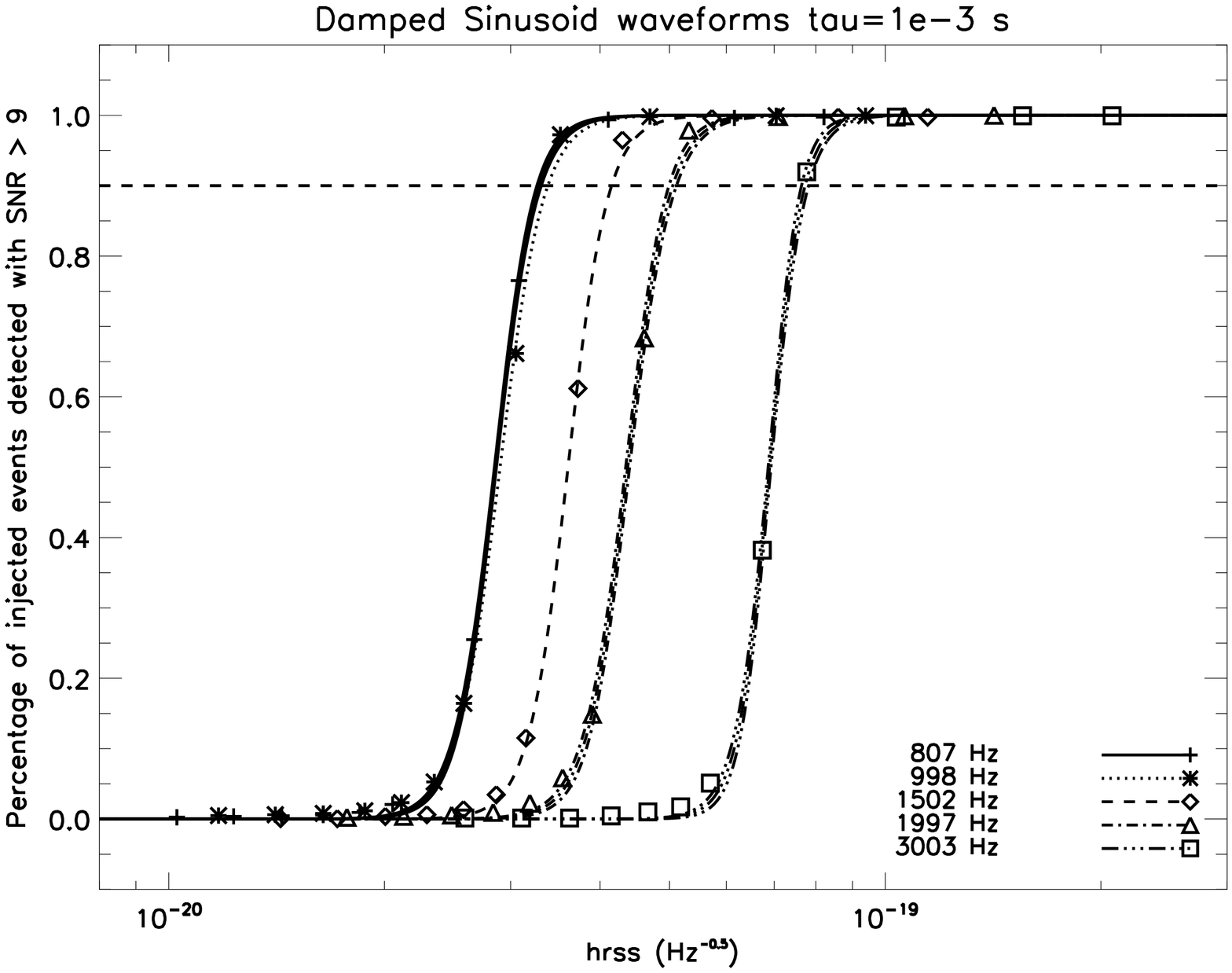}
		\caption[]{Similar to Fig. \ref{DS3}, efficiency curves for damped-sinusoid waveforms with $\tau=1$~ms.}
	\label{DS1}
	\end{center}
\end{figure}

In Figs. \ref{DS3}-\ref{DS1} we report the results obtained for damped-sinusoid waveforms with $\tau$ values of $0.3$~ms, $1$~ms and $1.5$~ms. For the given $h_{rss}$ and $f_0$ values, the detection efficiency decreases with decreasing $\tau$. On the other hand, for a given $\tau$ but different characteristic frequencies, the detection efficiency decreases with increasing $f_0$.

\begin{figure}
\begin{center}
		\includegraphics[width=6cm,angle=90]{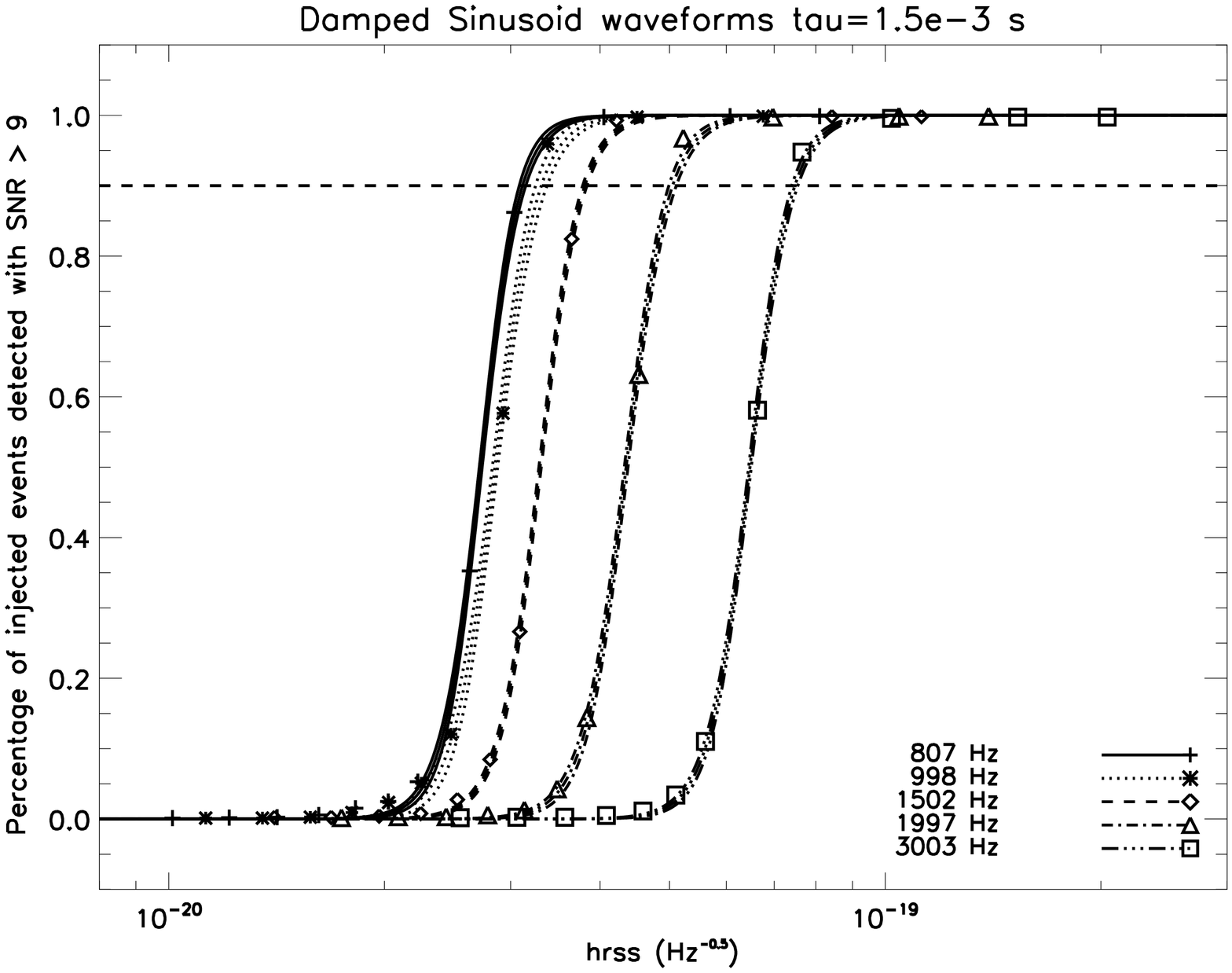}
		\caption[]{Similar to Figs. \ref{DS3}-\ref{DS1}, efficiency curves for damped-sinusoid waveforms with $\tau=1.5$~ms.}
	\label{DS1.5}
	\end{center}
\end{figure}

\begin{figure}
\begin{center}
		\includegraphics[width=6cm,angle=90]{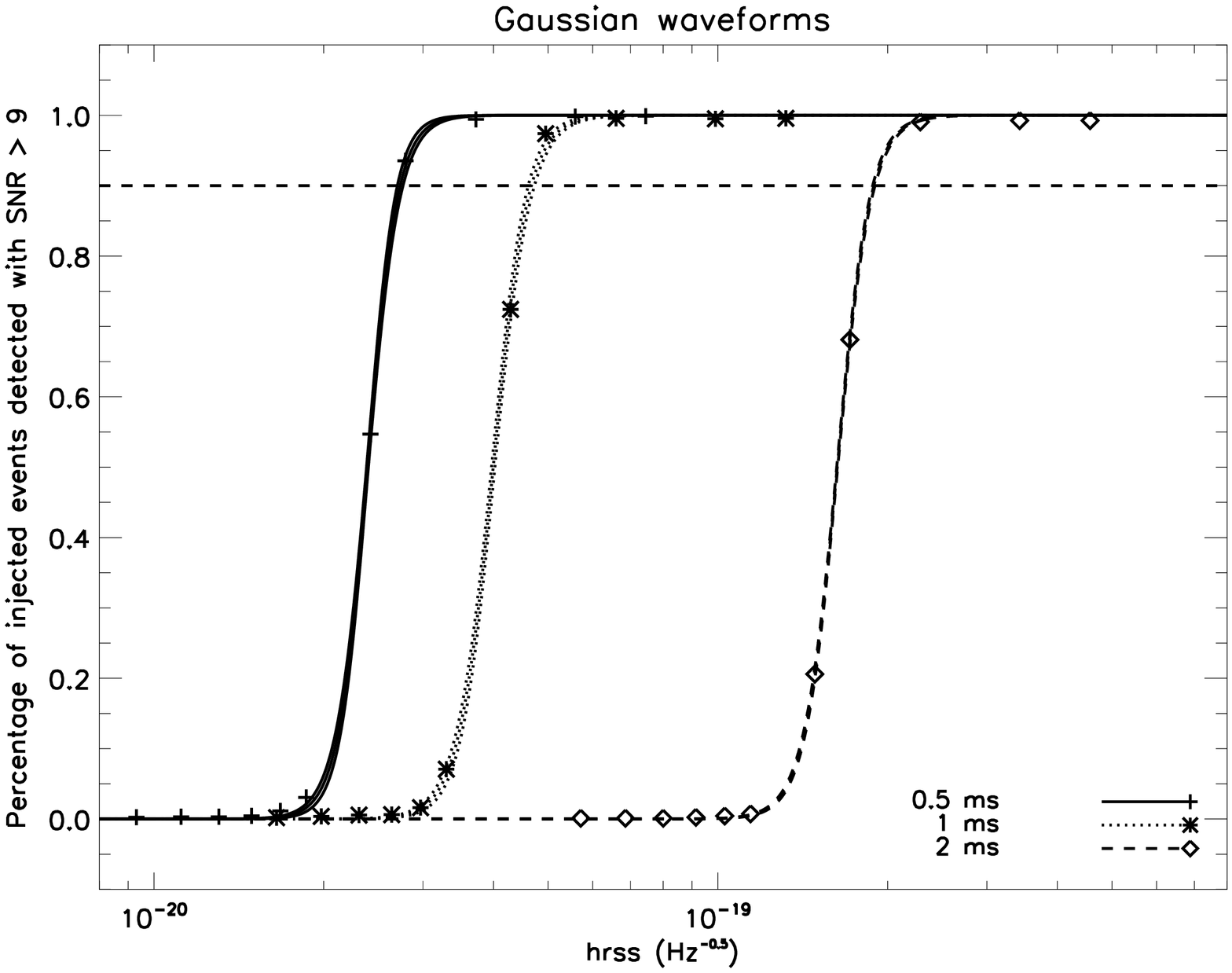}
		\caption[]{Efficiency curves for Gaussian waveforms with different $\sigma$, as a function of the injected $h_{rss}$ (see the caption of Fig. \ref{SGQ5eff} for further details).}
	\label{Geff}
	\end{center}
\end{figure}

Finally, in Fig. \ref{Geff} we show the results obtained for the simplest type of simulated waveforms, i.e. the Gaussian ones. The $5\sigma$ duration of these signals are between $1.5$~ms and $7.5$~ms. For a given $h_{rss}$ value, a higher $\sigma$ in the time domain implies that the energy of the Gaussian is in the low frequency region of the detection bandwidth, hence the detected $SNR$ (and thus the detection efficiency) is lower. This causes the detection efficiency to decrease for increasing $\sigma$. 

To estimate the $h_{rss}$ at $90\%$ pipeline efficiency for signals with $SNR>9$, we fit the points in Figs. \ref{SGQ5eff}-\ref{Geff} with sigmoid functions of the form:
\begin{equation}
	\epsilon=\frac{1}{1+\exp[-(\log h_{rss}-p_{1})/p_{2}]}
	\label{fit}
\end{equation}
The $90\%$ $h_{rss}$ values reported in tables \ref{table1}-\ref{table3} (third column) are obtained by setting in Eq. (\ref{fit}) $p_1$ and $p_2$ equal to their best fit values, and determining the $h_{rss}$ corresponding to the $90\%$ efficiency, $\epsilon^{*}$. The reported errors correspond to the $\pm2\sigma$ uncertainty on the best fit curve. 
We stress that the derived upper-limits (last column in tables \ref{table1}-\ref{table3}), are affected by a $+20\%-40\%$ systematic error, related to the uncertainties in the calibration of $h-$reconstruction (see also section \ref{Virgo}). The lowest $h_{rss}$ upper-limit is obtained for the sine-Gaussian waveform at frequency $f_0=497$~Hz with $Q=5$, for which $h^{SG}_{rss}\sim 2.09\times10^{-20}$~Hz$^{-1/2}$. 

We compare our results for sine-Gaussian waveforms with Q=5, with those obtained by \cite{Abbott030329} for sine-Gaussian waveforms with $Q=4.5$, in association with GRB~030329 during LIGO S2. The sensitivity of the Hanford detectors during S2 was similar to Virgo during C7. Moreover, the visibility of GRB~030329 from LIGO ($\sqrt{(F^{0}_{+})^{2}+(F^{0}_{\times})^{2}}=0.37$), was nearly equal to the one of GRB~050915a from Virgo ($0.38$). LIGO's lowest strain upper-limit, $h_{rss}=2.1\times10^{-20}$~Hz$^{-1/2}$ (note that according to the different definitions, the upper-limits reported in Table I of \cite{Abbott030329} should be divided for $\sqrt{\frac{(F^{0}_{+})^{2}+(F^{0}_{\times})^{2}}{2}}$ before comparing with the results reported in our Table \ref{table1}, where the quoted $h_{rss}$ strains do not contain the attenuation for the antenna pattern (see Eq. \ref{hrss})), was obtained at $f_0\sim 250$~Hz. In our case, we get the lowest value of $h_{rss}\sim 2.09\times10^{-20}$~Hz$^{-1/2}$ at $\sim 500$~Hz. At higher frequencies, around $1000$~Hz, the LIGO upper-limit is $h_{rss}=6.5\times10^{-20}$~Hz$^{-1/2}$, to be compared with $h_{rss}\sim 2.96\times10^{-20}$~Hz$^{-1/2}$ in the Virgo case. We stress that the LIGO procedure is based on the cross-correlation between the output of the two Hanford detectors, while our search is a single detector analysis. 

\begin{table}
\begin{center}
\caption{\label{table1}$h_{rss}$ upper-limits for sine-Gaussian waveforms (see Figs. \ref{SGQ5eff}-\ref{SGQ15eff}). The first two columns give details on the waveform parameter state, the third column is the $h_{rss}$ for which $90\%$ efficiency is reached in detecting simultaed signals at $SNR>9$. The error-bars account for the errors on the best fit values of the sigmoid function parameters. Note that these $h_{rss}$ values are affected by a systematic error, as described in Sec. \ref{Virgo} and Sec. \ref{deteff}. }
\vspace{0.5cm}
\begin{tabular}{ccc}
$Q$&$f_{0} $ & 90\% $h^{SG}_{rss}\times 10^{20}$ \\
& (Hz)& (Hz$^{-1/2}$)\\
\hline
5& 203 & $2.42\pm0.04$\\
5 & 497 & $2.09^{+0.02}_{-0.04}$\\
5 & 803 & $2.59^{+0.02}_{-0.01}$\\
5 & 1001 & $2.96\pm0.03$\\
5 & 1503 & $3.78^{+0.08}_{-0.07}$\\
15 & 203 & $3.34^{+0.06}_{-0.07}$\\
15 & 497 & $2.33^{+0.02}_{-0.03}$\\
15 & 803 & $2.79^{+0.05}_{-0.04}$\\
15 & 1001 & $2.68\pm0.03$\\
15 & 1503 & $3.04^{+0.01}_{-0.02}$\\
\end{tabular}
\end{center}
\end{table}

\begin{table}
\begin{center}
\caption{\label{table2}$h_{rss}$ upper-limits for damped-sinusoid waveforms (see Figs. \ref{DS3}-\ref{DS1}). The first two columns give details on the waveform. The last column gives the $h_{rss}$ for $90\%$ efficiency. Note that these $h_{rss}$ values are affected by a systematic error, as described in Sec. \ref{Virgo}  and Sec. \ref{deteff}.}
\vspace{0.5cm}
\begin{tabular}{ccc}
$\tau$ & $f_{0}$ & 90\% $h^{DS}_{rss}\times 10^{20}$ \\
(ms) & (Hz) & (Hz$^{-1/2}$)\\
\hline
0.3& 807 & $3.83^{+0.03}_{-0.02}$\\
0.3 & 998 & $4.09\pm0.06$\\
0.3 & 1502 & $5.03\pm0.08$\\
0.3 &  1997& $5.48\pm0.07$\\
0.3 & 3003& $8.0\pm0.1$\\
1 &  807& $3.29\pm0.03$\\
1 &  998& $3.39^{+0.05}_{-0.06}$\\
1 &  1502&$4.16^{+0.05}_{-0.06}$\\
1 &  1997& $5.06^{+0.07}_{-0.08}$\\
1& 3003 & $7.7^{+0.1}_{-0.1}$\\
1.5 & 807 & $3.12\pm0.04$\\
1.5 & 998 & $3.32^{+0.06}_{-0.07}$\\
1.5& 1502 & $3.81^{+0.04}_{-0.03}$\\
1.5 & 1997 & $5.05^{+0.08}_{-0.07}$\\
1.5 & 3003 & $7.48^{+0.10}_{-0.09}$\\
\end{tabular}
\end{center}
\end{table}

\begin{table}
\begin{center}
\caption{\label{table3}$h_{rss}$ upper-limits for Gaussian waveforms (see Fig. \ref{Geff}). The first column gives details on the waveform. The last column gives the $h_{rss}$ for $90\%$ efficiency. Note that these $h_{rss}$ values are affected by a systematic error, as described in Sec. \ref{Virgo}  and Sec. \ref{deteff}.}
\vspace{0.5cm}
\begin{tabular}{ccc}
$\sigma$ & 90\% $h^{G}_{rss}\times 10^{20}$ \\
(ms) & (Hz$^{-1/2}$)\\
\hline
0.5& $2.74^{+0.02}_{-0.04}$\\
1 & $4.68^{+0.07}_{-0.08}$\\
2 & $18.9^{+0.2}_{-0.1}$\\
\end{tabular}
\end{center}
\end{table}

\subsection{Astrophysical interpretation}

As described in section \ref{GW emission}, GWs could give a direct information on the GRB progenitor's identity. Of course, the critical aspect in theoretical models for the production of GWs in association with long GRBs, is the fraction of energy expected to be emitted in GWs, $E_{GW}$, during the phases when dynamical instabilities develop. 

Sources radiating energy $E_{GW}$ could produce an extremely small $h(t)$ signal at the detector, depending on the emission pattern. Nevertheless, we can always associate a strain $h(t)$ at the detector with some minimum amount of $E_{GW}$ radiated by the source, selecting an optimistic emission pattern. This is in fact the spirit of the analysis presented here, where attention was mostly devoted to that phases of GW emission dominated by a $l=m=2$ emission pattern (i.e. having a maximum along the line of sight). Considering that the redshift of GRB~050915a is not known, in what follows we will assume that this burst was at the distance of the nearest long GRB ever observed, i.e. GRB~980425 at $d_L\sim40$~Mpc. Moreover, we will focus attention on the $h_{rss}$ upper-limit obtained for the sine-Gaussian waveform at frequency $f_0=203$~Hz with $Q=5$. As shown in \ref{secondaappendice}, in this case the radiated energy is computed as:
\begin{eqnarray}
	E_{GW}\simeq (h^{SG}_{rss})^{2}\frac{ c^{3}d^{2}_{L}2 \pi^{2} f^{2}_0}{5 G (1+z)}
	\label{EUL}
\end{eqnarray}
which gives an energy upper-limit of:
\begin{equation}
	E^{UL}_{GW}\simeq 350 M_{\odot}(d_{L}/40 \rm{Mpc})^{2}
	\label{nultimo}
\end{equation}
When Virgo is running at its nominal sensitivity, the noise strain around $\sim 200$~Hz is expected to be about a factor of $15$ lower than during C7 (see Fig. \ref{Fig1}). Thus, if we assume to have a $SNR$ distribution confined below $SNR=9$, then the energy upper-limit given in Eq. (\ref{EUL}) would be lowered of a factor of $\sim 225$. Further improvement may also come in the case of optimal orientation: e.g. if GRB~050915a was optimally oriented with respect to the Virgo antenna pattern, the upper-limit in Eq. (\ref{nultimo}) would be a factor of $\left((F^{0}_{+})^{2}+(F^{0}_{\times})^{2}\right)^{-1}\simeq 7$ lower. Moreover, the joint collaboration with LIGO will help in setting upper-limits, since a coincidence search using three or four detectors will be a powerful tool in reducing the tail observed in the $SNR$ distribution of the on-source region.  

Some of the most optimistic predictions for the emission of GWs when instabilities develop in the rotating core of the massive GRB progenitor or in the disk surrounding the final BH, give an upper-limit estimate of the order of $\sim 0.1 M_{\odot}$ (e.g. \cite{Kobayashi2003} for the case of a merger of two blobs of $1~M_{\odot}$ each, formed in the fragmentation of a collapsing core). We thus conclude that, \textit{under the optimistic assumptions of optimal orientation and distance of $40$~Mpc}, the Virgo detector at its nominal sensitivity will start reaching the level of theoretical upper-limit estimates for GW emission by long GRB progenitors.

\section{Conclusion}
\label{conclusioni}
We have presented the first analysis of Virgo data in coincidence with a GRB trigger, aimed to search for a burst of GWs associated with the long GRB~050915a, occurred during Virgo C7 run. We have analyzed a time window of $180$~s around the GRB trigger time, and about $4.6$ hours of off-source data, corresponding to a single lock stretch. The result of this analysis is a set of loudest event upper-limits on the strain of an astrophysical GW signal occurring in association with GRB~050915a. The evaluation of the pipeline and detector efficiency for detecting signals showing up as events with $SNR$ above the loudest observed in the on-source region, was performed by adding a set of simulated burst-type signals to the off-source data, at randomly selected times. The waveforms of the simulated signals were chosen taking into account present uncertainties in the predictions for GW emission associated with GRB progenitors. linked with the ringing of the final BH.

The best upper-limit strain amplitudes obtained in our analysis are of the order of $h_{rss}=(2-4)\times10^{-20}$~Hz$^{-1/2}$ around $\sim 200-1500$~Hz, affected by a $\sim +20\%-40\%$ systematic error. On the basis of these results we conclude that, when running at nominal sensitivity, Virgo will start putting interesting astrophysical constraints for GW emission in association with GRBs at distances comparable to GRB~980425. 

Short GRBs, probably associated with the merger of compact binaries and occurring at lower redshifts \cite{BergerReviewShort,Nakar}, will represent even more promising targets. The procedure for the analysis presented here may in fact also be extended to the study of these sources, especially for the phases of merger and ring-down, but also for the last stages of the earlier in-spiral phase. The kind of search implemented here, while expected to be less efficient than a matched filtering approach, has the great advantage of avoiding strong dependence on exact knowledge of the in-spiral waveforms. Finally, these kinds of studies will be of great benefit for the joint collaboration with LIGO, in view of which is the hope that a coincident detection could occur with three or four interferometers, during the explosion of a relatively near GRB.

\subsection{Acknowledgments}The authors gratefully acknowledge the support of the Istituto Nazionale di Fisica Nucleare -INFN, of the Centre National de la Recherche Scentifique -CNRS, and of the European Gravitational Observatory -EGO. A. Corsi acknowledges the support of a VESF fellowship, funded by EGO and hosted at Istituto di Astrofisica Spaziale e Fisica Cosmica - IASF-Rome/INAF, on the project titled ``Gravitational Waves by Gamma-Ray Bursts''. A. C. thanks Luigi Piro for useful comments/suggestions and Pietro Ubertini for hosting the VESF project at IASF-Rome.

\appendix
\section{Details on the sine-Gaussian waveform choice}
\label{appendice}
In what follows, we address the question on how to mimic GW emission from collapse, fragmentation or bar instabilities. We review some results reported in the literature that are useful for our GRB analysis. 

Consider a source of gravitational radiation characterized by a mass quadrupole tensor $D_{i,j}$ \cite{LandauLifshitz1975}. The transverse-traceless components of the metric perturbation are related to the transverse-traceless components of the quadrupole tensor by the following relation:
\begin{equation}
	h^{TT}_{i,j}=\frac{2G}{3c^{4}d}d^{2}D^{TT}_{i,j}/dt^{2}
\end{equation}
where $i,j,k=1,2,3$. The waveforms of the radiation received by an observer at distance $d$ (much greater than the dimensions of the source) can be taken to be plane and depend on the relative (angular) orientation between the
observer and the source. In general, using a system of orthonormal spherical coordinates $(r,\theta,\phi)$ and considering an observer direction making an angle $\theta_{0}$ with the $x_3$-axis and $\phi_{0}$ with the $x_1$-axis, the two polarization states in the plane perpendicular to the direction of propagation can be characterized by the following expressions for the two non-vanishing components of
the perturbation to the galilean metric \cite{LandauLifshitz1975}:
\begin{equation}
h_{\times}= h_{\theta_{0}\phi_{0}} =-\frac{2G}{3c^{4}d}\ddot{D}_{\theta_{0}\phi_{0}}
\label{hxg} 
\end{equation}
\begin{equation}
h_{+}=h_{\theta_{0}\theta_{0}}=-h_{\phi_{0}\phi_{0}}=-\frac{G}{3c^{4}d}(\ddot{D}_{\theta_{0}\theta_{0}}-\ddot{D}_{\phi_{0}\phi_{0}})
\label{h+g}	
\end{equation}
where $D_{\phi_{0}\phi_{0}}$, $D_{\theta_{0}\phi_{0}}$, $D_{\phi_{0}\theta_{0}}$ are the physical components of $D_{i,j}$ projected along the directions of the spherical unit vectors. 

Consider now the particular case of a system characterized by a mass quadrupole tensor:
\begin{equation}
D_{\alpha,\beta}=
\left(\begin{array}{ccc}D_{11}&D_{12}&0\\D_{21}&D_{22}&0\\0&0&D_{33}\\
\end{array}\right)
\label{quadrupolo}
\end{equation}
with respect to a set of fixed inertial axes $(x_l, x_2, x_3)$, where $x_3$-direction could assume the invariant one of the angular momentum or the rotation. No such physical meaning is assigned to the $x_1$- and $x_2$-axes. A large class of realistic astrophysical systems turn out to have a mass quadrupole tensor of the form given in Eq. (\ref{quadrupolo}). These include, for example, binary systems, rotating ellipsoidal objects and pulsating/rotating ellipsoids \cite{BeltramiChau1985}. The latter case, for large amplitude pulsation, corresponds to explosion and collapse \cite{BeltramiChau1986,SaenzShapiro}. The mass quadrupole tensor in Eq. (\ref{quadrupolo}) is thus relevant for the case of a GRB progenitor, for which we expect the $x_{3}$-axis to be the rotational axis of the collapsing core (and of the subsequent BH plus accretion disk system), along which the GRB jet is launched. 

For such systems, using the freedom of rotation about the $x_3$-axis, so that a reference system can always be chosen to have the observer on the $x_1$-$x_3$ plane ($\phi_0=0)$, one has:
\begin{equation}
h_{\times}=-\frac{2G}{3c^{4}d}\ddot{D}_{12}\cos\theta_{0}
\label{hx} 
\end{equation}
\begin{equation}
h_{+}=-\frac{G}{3c^{4}d}\left[(\ddot{D}_{11}-\ddot{D}_{22})+(\ddot{D}_{33}-\ddot{D}_{11})\sin^{2}\theta_{0}\right]
\label{h+}	
\end{equation}
 
Using the simplified assumption of a rigid, uniform, ellipsoid rotating with an angular velocity $\omega$ around the $x_3$-axis, the quadrupole mass tensor has a time-independent expression in the frame $x'_1,x'_2,x'_3$ co-moving with the rigid rotating object, where non-diagonal elements are null (due to the reflection symmetry of the mass distribution). Using the coordinate transformations from one system to another, one can express the quadrupole tensor components in the $x_1,x_2,x_3$ coordinate system as a function of the time-invariant $D'_{\alpha,\beta}$ ones in the $x'_1,x'_2,x'_3$ system:
\begin{eqnarray}
\nonumber D_{11}=D'_{11}\cos^{2}(\omega t)+D'_{22}\sin^{2}(\omega t)\\
D_{12}=D_{21}=\frac{1}{2}\sin(2\omega t)(D'_{11}-D'_{22})\label{comovD}\label{nulli}\\
\nonumber D_{22}=D'_{11}\sin^{2}(\omega t)+D'_{22}\cos^{2}(\omega t)\\
\nonumber D_{33}=D'_{33}
\end{eqnarray}
From these expressions it is evident that $\ddot{D_{11}}=-\ddot{D_{22}}$ and $\ddot{D_{33}}=0$, so that
substituting in equations (\ref{hx}) and (\ref{h+}) one gets:
\begin{equation}
h_{\times}=
\frac{4G\omega^{2}}{3c^{4}d}(D'_{11}-D'_{22})\sin(2\omega t)\cos\theta_{0}
\label{hxfinale}
\end{equation}
\begin{equation}
h_{+}=
\frac{1}{2}\frac{4G\omega^{2}}{3c^{4}d}(D'_{11}-D'_{22})\cos(2\omega t)\left(1+\cos^{2}\theta_{0}\right)
\label{h+finale}
\end{equation}

Considering that we expect to be observing the GRB on-axis ($\theta_0=0$ in (\ref{hxfinale}) and (\ref{h+finale})), the signal is circularly polarized. Equations (\ref{hxfinale}) and (\ref{h+finale}) do apply also to the case of a binary system or to a bar-like structure, which are all thought to play a role in GRB progenitors \cite{Kobpolarization}. 

An equivalent but useful way to expand the waveforms is in terms of $l=2$ pure-spin tensor harmonics. For a transverse, traceless tensor, in the quadrupole approximation, the only components that can enter are the basis states usually labeled by $T^{E2,lm}$ \cite{kochanek}. The expansion of the GW (transverse-traceless component of the metric perturbation) in this basis can be written as:
\begin{equation}
	 h^{TT}_{ij}=-\frac{2G}{c^{4}d}A_{2m}T^{E2,2m}
	 \label{scomposizione}
\end{equation}
where there is an implicit summation over $m$. Using the explicit representation of these harmonics in the orthonormal spherical coordinates (see \cite{kochanek} for details), it is possible to show that: (i) in the case of a mass distribution having a quadrupole mass tensor of the form (\ref{quadrupolo}), which corresponds to a system with an x-y plane reflection symmetry (i.e. $D_{13}=D_{23}=0$), one has $A_{2\pm1}=0$; (ii) for a rigidly rotating ellipsoid, since equations (\ref{nulli}) are valid (i.e. $\ddot{D_{11}}=-\ddot{D_{22}}$ and $\ddot{D_{33}}=0$), one has $A_{2\pm0}=0$. Thus, we can say that GW emission from binary systems or rotating rigid ellipsoids are dominated by the $l=\left|m\right|=2$ mode, for which the wave amplitude is maximized along the rotational axis.
\\\newline

\section{Energy radiated in GWs}
\label{secondaappendice}
In what follows we give details on the procedure followed to determine the energy upper-limit for the sine-Gaussian waveform with $Q=5$ at $\sim 200$~Hz.

The energy radiated in GWs is computed as:
\begin{equation}
E_{GW}=\frac{c^{3}d^{2}_{L}}{16\pi G}\int d\Omega \int^{+\infty}_{-\infty}(\dot{h}^{2}_+(t)+\dot{h}^{2}_{\times}(t))\frac{dt}{1+z}
\end{equation}
where the integration over the solid angle should be performed while considering the emission pattern. If the signal power at the detectors is dominated by a frequency $f_0$, as is the case for the sine-Gaussian waveforms, the above formula is approximated as:
\begin{equation}
E_{GW}\simeq\frac{c^{3}d^{2}_{L}}{16\pi G}(4\pi^{2} f^{2}_0)\int d\Omega \int^{+\infty}_{-\infty}(h^{2}_+(t)+h^{2}_{\times}(t))\frac{dt}{1+z}
\end{equation}
On the basis of Eqs. (\ref{hxfinale})-(\ref{h+finale}), we have:
\begin{eqnarray}
	\nonumber E_{GW}\simeq\frac{ c^{3}d^{2}_{L}2\pi^{2} f^{2}_0}{4 G (1+z)}\int^{1}_{-1} d(\cos\theta)\int^{+\infty}_{-\infty}dt\times\\ \times\left[\frac{1}{4}(1+\cos^{2}\theta)^{2}h^{2}_{+,0}(t)+\cos^{2}\theta h^{2}_{\times,0}(t)\right]\\
\end{eqnarray}
where $h_{+,0}(t)$ and $h_{\times,0}(t)$ are given by the plus and cross components in Eq. (\ref{piu})-(\ref{per}). Taking into account that $\int^{+\infty}_{-\infty}h^{2}_{\times,0}dt=\int^{+\infty}_{-\infty}h^{2}_{+,0}dt$, we can write:
\begin{eqnarray}
	\nonumber E_{GW}\simeq\frac{ c^{3}d^{2}_{L}2\pi^{2} f^{2}_0}{4 G (1+z)}\int^{1}_{-1} d(\cos\theta)\left[\frac{1}{4}(1+\cos^{2}\theta)^{2}+\cos^{2}\theta\right] \\\nonumber\int^{+\infty}_{-\infty}dt~ h^{2}_{+,0}(t)\\
\end{eqnarray}
The time integral is equal to $(h^{SG}_{rss})^{2}/2$, where $h^{SG}_{rss}$ has the values quoted in Table \ref{table1}. Thus we write:
\begin{eqnarray}
	\nonumber E_{GW}\simeq (h^{SG}_{rss})^{2}\frac{ c^{3}d^{2}_{L}2 \pi^{2} f^{2}_0}{5 G (1+z)}
	\label{EULapp}
\end{eqnarray}

\vspace{0.5cm}
\bibliographystyle{unsrt}
\bibliography{Corsi_GRB050915a}

\end{document}